\documentclass[journal]{IEEEtran}
\usepackage{cite}
\usepackage{amsmath,amssymb,amsfonts}
\usepackage{algorithmic}
\usepackage{graphicx}
\usepackage{textcomp}
\usepackage{xcolor}
\def\BibTeX{{\rm B\kern-.05em{\sc i\kern-.025em b}\kern-.08em
    T\kern-.1667em\lower.7ex\hbox{E}\kern-.125emX}}

\usepackage{orcidlink}
\usepackage{circledsteps}
\usepackage{enumitem}
\usepackage{hyphenat}
\usepackage{csquotes}
\usepackage{doi}
\newcommand{\thematicbreak}{\par\medskip}

\newcommand\helcircled[2][]{\ifmmode
\Circled[fill color=black,inner color=white,#1]{\mathsf{#2}}
\else
\Circled[fill color=black,inner color=white,#1]{\sffamily#2}
\fi
}

\renewcommand{\mkbegdispquote}[2]{\itshape\openautoquote}

\begin{document}

\title{Helveg: Diagrams for Software Documentation}
\author{%
    Adam~Štěpánek\orcidlink{0009-0008-9388-2546},\,%
    David~Kuťák\orcidlink{0000-0002-4346-6850},\,%
    Barbora~Kozlíková\orcidlink{0000-0003-0045-0872}, and\,%
    Jan~Byška\orcidlink{0000-0001-9483-7562}
    \thanks{All authors are with Masaryk University, Brno, Czech Republic.}
    \thanks{Jan Byška is also with University of Bergen, Bergen, Norway.}
}

\maketitle
\begin{abstract}

% === ORIGINAL ABSTRACT ===
% Getting acquainted with a large codebase can be a daunting task for software developers, both new and seasoned.
% The description of a codebase and its development should be the purpose of its documentation.
% However, software documentation, if it exists at all, is usually textual and accompanied only by simple static diagrams.
% It is also time-consuming to maintain manually.
% Even an API reference, which can be generated automatically from the codebase itself, has many drawbacks.
% It is limited to what it can extract from the codebase, is cumbersome to navigate, and fails to capture the interwoven nature of code.
% We explore an alternative approach centered around a node-link diagram representing the structure of a codebase.
% The diagram is interactive and filterable, providing details on demand.
% It is designed for automation, relying on static analysis of the codebase, and thus produces results quickly and offers a viable alternative to missing or outdated documentation.
% To evaluate this approach, we implemented a prototype named Helveg that is able to analyze and visualize C\# code.
% Testing with five professional programmers provided feedback on the approach's benefits and challenges, which we discuss in detail.

Software developers often have to gain an understanding of a codebase.
Be it programmers getting onboarded onto a team project or, for example, developers striving to grasp an external open-source library.
In either case, they frequently turn to the project's documentation.
However, documentation in its traditional textual form is ill-suited for this kind of high-level exploratory analysis, since it is immutable from the readers' perspective and thus forces them to follow a predefined path.
We have designed an approach bringing aspects of software architecture visualization to API reference documentation.
It utilizes a highly interactive node-link diagram with expressive node glyphs and flexible filtering capabilities, providing a high-level overview of the codebase as well as details on demand.
To test our design, we have implemented a prototype named Helveg, capable of automatically generating diagrams of C\# codebases.
User testing of Helveg confirmed its potential, but it also revealed problems with the readability, intuitiveness, and user experience of our tool.
Therefore, in this paper, which is an extended version of our VISSOFT paper with DOI \href{https://doi.org/10.1109/VISSOFT64034.2024.00012}{10.1109/VISSOFT64034.2024.00012}, we address many of these problems through major changes to the glyph design, means of interaction, and user interface of the tool.
To assess the improvements, this new version of Helveg was evaluated again with the same group of participants as the previous version.

\end{abstract}


%-------------------------------------------------------------------------

\section{Introduction}

% Why is documentation important?
% What does documentation look like?
% There isn't enough documentation. Why?
% What is API reference?
% Why is API reference not good enough?
% How is visualization used in docs?
% How are we trying to improve it?
% What is Helveg?
% What is the contribution of this paper?

Nearly every piece of software eventually reaches a point when it should be documented.
This is because technical documentation is the most sought-after learning resource~\cite{so_survey}.
Well-written documentation is the hallmark of good open-source projects and can be critical for their adoption~\cite{good_docs}.
Unfortunately, real-world documentation is often incomplete or confusing~\cite{oss_survey} due to time constraints.
% Docs are important because people want to learn from them. Unfortunately, docs are often less than great.

As of 2022, software documentation is typically written in a descriptive or structured textual form and is hosted on the web to be easily accessible~\cite{so_survey, tech-writing}.
The text is frequently enhanced with code snippets and less often with simple static charts and diagrams~\cite{doc-strategies}.
Developers often use automatic generators to extract special comments and metadata directly from the source code, reducing the hassle of writing documentation.
These generators, such as Doxygen~\cite{doxygen}, Sphinx~\cite{sphinx}, or DocFx~\cite{docfx}, typically output a page or an entry for every significant code element (e.g., a type or a function).
The documentation generated this way is often %collectively
described as an \textit{API reference}.
% The state of documentation writing these days.

An API reference is used by developers in broadly two ways~\cite{api_docs_usage}.
They may rely on it to systematically build their internal high-level mental model of the API, its capabilities, and its nomenclature.
They might also opportunistically use it to learn the low-level details of a specific API member they are attempting to use.
The main advantage is its automatic creation, which saves precious development time and makes it easier to keep documentation up-to-date with the code.
However, an API reference has many drawbacks.
It is limited to code metadata and the contents of \textit{documentation comments}.
These comments are written manually by the developers and, thus, are often omitted.
The navigation of an API reference usually relies on hyperlinks, full-text search, and a table of contents mimicking the module hierarchy.
Thus, this limited navigability makes it cumbersome to look for specific arrangements of code elements like types bound together through composition.
Therefore, authors often use multiple writing methods in practice, combining their advantages.
They write manually about key concepts spanning multiple code elements and delegate the details of each code element to the automatic API reference generators.
% These days documentation is written half manually, half automatically.

While documentation can benefit greatly from using visualization~\cite{whiteboard, rost_survey}, its use is currently limited.
It is usually confined to diagrams that describe the project's architecture, explain its type hierarchies, or compare the project to its competitors.
While these visualizations are undoubtedly helpful, they are seldom interactive, are often created manually, and usually act only as a support for their surrounding text.
Therefore, they make writing documentation even more demanding because authors need knowledge of a graphic editor or a tool like GraphViz~\cite{graphviz}.
Even diagrams, such as those generated by Doxygen~\cite{doxygen}, are only static images and thus do not utilize the interactive potential of digital platforms.
% Vis in docs are rare and mostly static.

To see the challenges with software documentation, consider a scenario where a new developer is assigned to a project. 
Navigating the existing codebase can be challenging without well-maintained documentation or the help of a senior colleague---both of which are frequently unavailable.
Often, the newcomers are left only with an API reference that offers a detailed view of each code piece but not a high-level overview of the entire project.
As the newcomers are unfamiliar with the codebase, they cannot effectively utilize full-text search.
Consequently, they are left to browse the entire documentation, hoping to stumble upon the information they need.
This process is inefficient and often frustrating.
Alternatively, they can generate a static diagram showing the whole project.
However, these diagrams typically result in unreadable hairballs or, conversely, omit critical details.
% There is a place for something more than text and static vis...

In this paper, we propose a visualization allowing systematic exploration of an API much like an API reference.
In our approach, the documentation is still generated automatically, but is navigated through an interactive node-link diagram.
This diagram enables a high-level, exploratory analysis of a codebase without trudging through its implementation details.
However, it is also flexible enough to provide these details on demand through filtering and interaction.
% ...an interactive form of an API reference.

\begin{figure}[tb]
  \centering
  \includegraphics[width=1.0\linewidth]{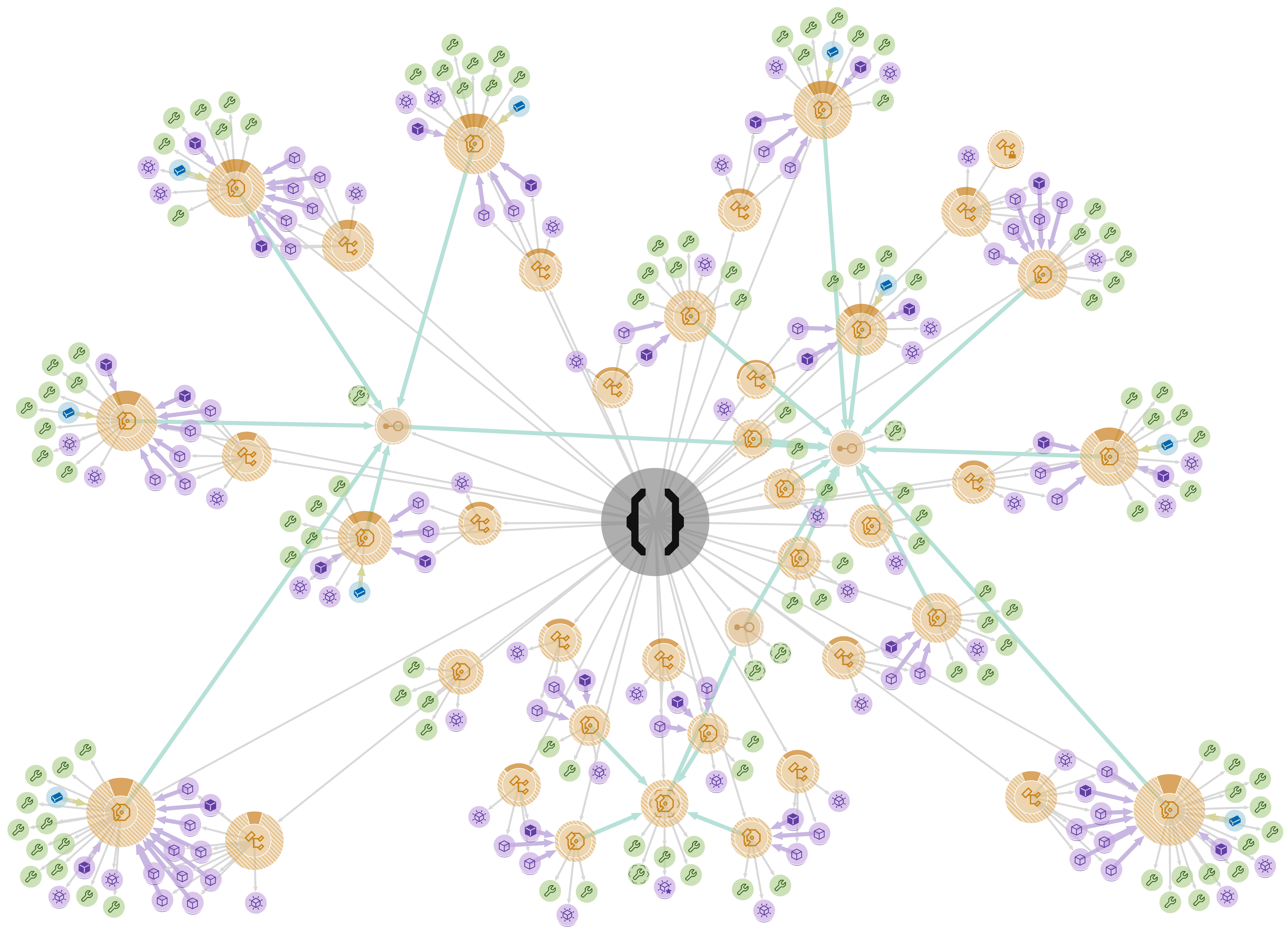}
  \caption{
    A sample diagram of a namespace with types (yellow nodes), methods (purple nodes), properties (green nodes), and fields (blue nodes).
    It contains several relations: declaration hierarchy (gray edges), type inheritance (teal edges), and method return types (purple edges). Labels were omitted.
    }
  \label{fig:example} 
\end{figure}

To test this concept, we built a prototype tool called Helveg.
Helveg can analyze a codebase written in the C\# programming language and automatically output a web application with an interactive diagram, representing the codebase.
Figure~\ref{fig:example} contains one such diagram, the nodes of which represent C\# entities, for example types, with links connecting those that are related, for instance through type inheritance.
Although Helveg provides a general overview of a codebase, we designed it to function mainly in place of or alongside a web-hosted API reference, and, in this paper, we focus on this use case.
% We built a prototype.

This paper is an extended version of a work we presented at VISSOFT~2024~\cite{op}.
It enhances the design and interactivity of the diagram while also addressing the major shortcomings identified by the participants of the user study as described in the original paper.
To validate this claim, we conducted a second round of testing with the developers, who gave their feedback on the tool's previous version.

In summary, the original paper contributes with:

\begin{itemize}
    \item An interactive diagram designed to serve as or complement an API reference when exploring codebases.
    \item An open-source prototype implementation of this design, which is available at \url{https://gitlab.com/helveg/helveg}.
    \item A preliminary user study of the concept with professional C\# developers.
\end{itemize}

The contributions of this extended version also include:

\begin{itemize}
    \item New and simpler glyph design with custom icons, revised visual components, improved color schemes, and support for more C\# language features.
    \item Many changes to the tool's interactivity and intuitiveness, such as mouse and keyboard controls, a new filtering mode, a tutorial, and a tree view, which resembles typical API reference navigation.
    \item Second user study with the same group of developers, evaluating the effectiveness of the changes.
\end{itemize}

%-------------------------------------------------------------------------

\section{Background and Related Work}

% Background: What should the reader know before reading the rest of the paper?
% Related Work: What else exists and does something similar? Who made it?
% How does vis in docs relate to the rest of softvis?
% Is it connected to program comprehension?

Documentation is connected to the topic of program comprehension, which many software visualization (SV) methods aim to improve.
These methods often focus on visualizing the structure of a program's source code, the program's behavior at runtime, or its evolution.
Since this paper proposes a documentation approach that concerns itself only with the static structure of a codebase, we further delve into techniques that do the same.
We also compare many existing open-source and commercial programs that enhance software documentation or visualize code structure.
Regarding SV as a whole, we refer to literature reviews by Chotisarn et al.~\cite{slr} and Bedu et al.~\cite{slr_2}.

\subsection{Visualization of Code Structure}

% What other research has been done and by whom?
% How are they different from one another?
% What are code maps? What are their pros and cons?
% What are code cities? What are their pros and cons?
% What abstract 2D approaches exist? What are their pros and cons?

A simple way of visualizing a large amount of source code is to scale it down.
This is how, in the code-map metaphor~\cite{code_map_review}, code lines become colored, elongated rectangles.
It can be used to visualize length, age, programming language, and other code file metrics.
SeeSoft by Eick et al.~\cite{seesoft} is a well-known example of this method.
This technique continues to be researched in recent years, as proven by CodePanorama by Etter and Mehta~\cite{codepanorama}, and it is even included in the popular Visual Studio Code editor as its Minimap feature~\cite{vscode_minimap}.

A wholly different class of code structure SV methods is based on real-world metaphors.
One such metaphor is the software city, in which codebases transform into 3D cities, where buildings represent units of code, such as classes, and their dimensions and colors correspond to code metrics.
The idea has given rise to diverse implementations such as Software World~\cite{software_world}, CodeCity~\cite{code_city}, CodeMetropolis~\cite{codemetropolis}, and BabiaXR~\cite{babiaxr}.
The city metaphor is further extended to islands populated by cities in IslandViz by Misiak et~al.~\cite{islandviz}.
While the metaphors are intuitive, they frequently omit relationships between code elements~\cite{city_metaphor_overview} and suffer from occlusion issues and perspective distortion common to all 3D visualizations~\cite{no_unjustified_3d}.

Apart from code maps and software cities, there is a plethora of 2D charts and diagrams.
These are designed for diverse purposes, spanning from visualization of software architecture and dependencies between modules to reverse engineering and optimization.
To name a few, Lanza and Ducasse~\cite{polymetric_views} presented polymetric views, a system with multiple views capable of displaying diverse metrics.
Gutwenger et al.~\cite{uml_approach} displayed UML class diagrams in a new way involving the minimization of crossings, orthogonal edges, uniform edge direction within type hierarchies, and other desirable criteria.
In E-Quality by Erdemir et al.~\cite{equality}, the codebase is displayed as a node-link diagram, where nodes are glyphs representing quality metrics.
Müller and Zeckzer~\cite{recursive_disk} also take a glyph-based approach, in which the structure and relations in a codebase become a circular treemap.
There are also methods utilizing rectangular and circular treemaps, like Git-Truck~\cite{git-truck} by Højelse et al.
Our visualization combines many aspects of the existing approaches.
It is also a node-link diagram with glyphs.
However, we designed it around the documentation use case: from basing our glyphs on existing iconography to tailoring the user interface to API reference users.

A common problem all of these approaches must tackle is scalability~\cite{slr_2}.
Even relatively small software projects involve a large amount of data.
Thus, visualizations must be equipped with some kind of filtering or aggregating mechanism to be readable.
The problem with scalability also applies to performance, as SV tool makers have to consider the complexity of their analysis and rendering algorithms.

To summarize, SV research in the area of code structure and software architecture offers a diverse group of techniques.
Our attempt shares some of the challenges with these techniques.
The scalability issue, in particular, is one we encounter as well.
However, we also face additional challenges stemming from our goal to offer experience comparable to an API reference.

\subsection{Existing Tools}
\label{sec:existing-tools}

% Are there any existing tools? Yes.
% What features do they have? Standards, Code Analysis, Interactivity, Portability, Diffing.
% How are they grouped?
% What is code analysis anyway?
% What drawbacks do they have?
% Where can I see more of them? Supplementary material.

%   1. Standards: UML, C4. Pseudo-standards: GraphViz & PlantUML formats.
%   2. Code Analysis / Automatic Generation.
%   3. Interactivity.   
%   4. Portability / docs-integration.
%       - Inclusion of documentation strings.
%       - Openable in a browser.
%       - Generatable using a command-line utility on a server without an IDE.
%   5. Diffing.
% - Tool groups
%   - code quality analyzers: *Depend, 
%   - UML class diagram generators: the JetBrains plugin, Visual Paradigm, Codartis, PlantUmlClassDiagramGenerator, pyp2uml
%   - non-UML code map generators: VS Code Map, Doxygen
%   - dependency visualizers: DependenSee, Dependency Cruiser
%   - text-to-diagram tools: GraphViz, PlantUML, Swimm, Structurizr, D2

Many existing tools can be used to visualize the structure of source code or software architecture.
In this section, we mention several examples and group them based on features relevant for use in software documentation.
For a more extensive comparison, see Supplementary Material~\cite{sm}.

\textbf{Conformance to standards.}
Architecture-visualizing tools typically produce diagrams that adhere to an existing standard, most commonly to UML class diagrams~\cite{uml}.
The advantage of this approach is that users are often already familiar with how the data is visualized.
Many tools can generate UML class diagrams from source code, such as the UML Diagrams plugin~\cite{jetbrains_uml} for JetBrains IDEs and Visual Paradigm~\cite{visual_paradigm}.
Apart from UML, there is also the less strict C4~model~\cite{c4} along with the Structurizr tool~\cite{structurizr}.
However, both UML and C4 were designed with static images in mind that can be, for instance, drawn on a whiteboard or printed.
Therefore, these tools usually do not take advantage of the potential for interactivity present in digital media, like the ability to expand and collapse nodes or provide detailed metadata on demand.
On the other hand, numerous architectural visualizers do not follow any standard, allowing them to include more language-specific details.
These include tools like the Visual Studio Code Map~\cite{code_map} (unrelated to the code-map metaphor) and all tools employing the city metaphor or a similar 3D approach.

\textbf{Code analysis / Automation.}
A significant distinguishing factor between the tools is whether they perform any \textit{code analysis}.
In the context of this paper, we define \textit{code analysis} as any data mining or processing step involving language parsing and semantic analysis.
For example, tools like Structurizr~\cite{structurizr} and D2~\cite{d2} are text-to-diagram utilities, turning a manually written description into a corresponding diagram.
On the other hand, NDepend~\cite{ndepend} and JArchitect~\cite{jarchitect} measure code quality and thus, apart from visualizing a codebase, also heavily analyze it.
Similarly, API reference generators, such as Doxygen~\cite{doxygen}, Sphinx~\cite{sphinx}, and Docfx~\cite{docfx}, can analyze a software project and generate static diagrams depicting inheritance hierarchies and attach them to relevant pages.
From the user's perspective, code analysis is the difference between describing a codebase manually and simply passing a path to a tool as an argument.
Tools performing code analysis can run automatically, which is a key requirement for serving as API references, since those are generated automatically on demand.

\textbf{Interactivity.}
While a static diagram of a small codebase can be readable, it often becomes cluttered as the codebase grows.
This issue can be largely avoided by providing details-on-demand features---letting the user interact with the visualization and choose what they want to focus on~\cite{shneiderman}.
For instance, the Emerge tool~\cite{emerge} features context menus, filtering, highlighting, and a force-directed layout algorithm.
However, Emerge focuses more on code quality than documentation and, for example, disregards documentation comments.

\textbf{Web support.}
Any alternative to an API reference must be ready for the web environment.
This is because software documentation tends to be hosted online, where anyone can access it.
Although many tools are interactive and can generate diagrams from data obtained by code analysis, they are distributed as binary programs with a graphical interface.
Therefore, they lose their interactivity when their diagrams are exported as static images.
These tools also often cannot run during automated builds, making them difficult to use along or in place of an API reference.
Commercial web platforms, such as CodeSee~\cite{codesee} and Swimm~\cite{swimm}, offer a web experience, but they focus on developer collaboration and maintainability rather than on code analysis.
There are also open-source tools like Emerge~\cite{emerge} and DependenSee~\cite{dependensee} that output their diagrams as self-contained web applications, which can be placed next to the documentation as is.
However, while they demonstrate their strength in embeddability, the scope of their analysis is limited to module dependencies.
The final mention goes to the Obsidian note-taking tool---a program for managing Markdown notes connected by hyperlinks~\cite{obsidian}.
It also features a node-link diagram displaying the connections between files.
Despite not performing code analysis, the Obsidian development team utilized it for their API reference.

\thematicbreak

In summary, none of the mentioned tools provides a combination of features that would let it improve upon the exploratory capabilities of an API reference.
They either cannot be automated, do not analyze the code sufficiently, are not interactive, or cannot be hosted on the web.

%-------------------------------------------------------------------------

\section{Requirement Analysis}
\label{sec:requirements}

This paper explores the assumption that interactive visualization of a codebase's structure can improve documentation and facilitate exploratory analysis.
Our solution was designed around this assumption while considering several requirements drawn from three sources.
The first source was a survey among professional programmers, programming tutors, and computer science students conducted within the scope of a master thesis~\cite{mgr}.
Another source was our comparison of existing tools and their strengths and weaknesses, which we summarized in the previous section.
Based on the information collected so far, we settled on the following goals.

\begin{enumerate}[series=req,label={\textbf{R\arabic*:}}, ref=R\arabic*, leftmargin=*]
    \item\label{req:familiar}\label{req:first} Output a visualization that feels familiar.
    \item\label{req:analyze} Generate the visualization automatically using data from source code analysis.
    \item\label{req:interact} Allow user interactions to enable exploration and filtering of the codebase on different levels of granularity.
    \item\label{req:embed} Output visualization in a format that can be embedded into a website.
\end{enumerate}

The final source of requirements was the feedback gathered from the first round of testing~\cite{op}.
We have decided to focus on glyphs, filtering, and interactivity rather than scalability, since improving the diagram's scalability would require us to build a brand-new prototype.
A complete rewrite of the tool would also create uncertainty.
Testing the changes to glyph design and the user interface in isolation from far-reaching changes stemming from a rewrite (e.g., a new layout algorithm) allows us to assess their advantages.
In the end, we decided that the new version of the prototype should adhere to the following requirements as well.

\begin{enumerate}[resume*=req]
    \item\label{req:replace-outlines} The glyph design should be intuitive and should use common techniques from visualization and graphic design, if possible, rather than irregular visualization ideas.
    \item\label{req:indicators} Glyphs should reflect not only the represented entity but also the status of its descendants. For example, it should be possible to tell when a node's subtree contains a compiler diagnostic.
    \item\label{req:ergonomic-filtering}\label{req:last} Filtering should be not only flexible but also easy and comfortable to use.
\end{enumerate}

Apart from the high-level requirements listed above, we also had to decide on the technical aspects of the implementation, as we needed to test the concept in practice.
In particular, we needed to consider what programming languages to support, as this decision is important for data mining algorithms.
In this regard, we identified C\# as an ideal target language for our prototype implementation.
As our main goal was to test the visual representations, we decided to support only one programming language.
The advantages of C\# are its large user base~\cite{so_survey}, numerous projects to visualize, and an API for directly inspecting the compiler's semantic model.
However, we are convinced that the concept presented in this paper also applies to other object-oriented programming languages.
The following section summarizes the specifics of the C\# programming language, whose knowledge is crucial for understanding our solution and its components.

%-------------------------------------------------------------------------

\section{C\# Language and its Specifics}
\label{sec:csharp}

% Why do we explain C#? Terminology, guarantees, and constraints
% What do we build upon? .NET Roslyn, NuGet, MSBuild
% What are solutions and projects?
% What is code analysis? Definovat code analysis jako to, co dělá Roslyn.
% What are types (class, struct, interface, delegates)?
% What are type members (field, property, methods, events)?
% What is `static`? No free functions -> statické třídy jsou jen containery, takže mají úplně jiný význam
% What is the entity hierarchy? Solution vs project vs assembly vs namespace
% What is entity kind and type kind?
% What guarantees do we have? NuGet Packages nemohou dělat cykly

Our visualization design assumes familiarity with object-oriented programming (OOP) and particularly the C\# programming language.
This section explains several C\# terms, constraints, and guarantees relevant to this paper.

\textbf{The .NET platform.}
The C\# language is a part of the .NET platform, which envelops several other languages, runtimes, and related tooling.
There are three relevant pieces of tooling: Roslyn, NuGet, and MSBuild.
The .NET Compiler Platform (``Roslyn'')~\cite{roslyn} is the C\# compiler, providing \textit{code analysis} APIs, which allow others to build tools that understand C\#'s syntax and semantics.
NuGet~\cite{nuget} is .NET's package manager facilitating the resolution and acquisition of external dependencies.
Finally, Microsoft Build Engine (MSBuild)~\cite{msbuild} is the build system tying Roslyn and NuGet into a pipeline that transforms a C\# project, its source code, and dependencies into a binary artifact called \textit{an assembly}.
% Finally, Microsoft Build Engine (MSBuild)~\cite{msbuild} ties Roslyn and NuGet into a pipeline that transforms C\# code and its dependencies into a binary artifact called \textit{an assembly}.

\textbf{The project system.}
MSBuild works with \textit{projects} and \textit{solutions}.
Projects define what NuGet packages are downloaded, what C\# files are fed to Roslyn, and any other steps performed throughout the build process.
Visual Studio Solutions are then simply sets of projects.
MSBuild and NuGet prevent projects and packages from having circular dependencies.

\textbf{The type system.}
Object classes are the core concept of OOP.
In C\#, classes are just one kind of \textit{types} the language has to offer.
There are five \textit{type kinds} in total: classes, structures, enumerations, interfaces, and delegates.
All types must be declared inside a hierarchical \textit{namespace}.
The top-level namespace is the \textit{global namespace}.
Types may contain \textit{type members}, which fall into one of four categories: fields, methods, properties, and events.
Both types and their members have an \textit{accessibility}---a degree to which other pieces of code can interact with them.
For example, \textit{public} methods can be called from anywhere, while \textit{private} ones are accessible only within their declaring type.
Classes and structs can also be marked as \texttt{record}s, designating them as types primarily for data encapsulation.
For a more thorough description of C\# types, see .NET's Common Type System~\cite{dotnet_cts}.

\textbf{The static modifier.}
Class types and type members can have the \texttt{static} modifier.
This modifier has special importance since \texttt{static} members are accessible without an instance of their declaring type.
Static classes are then types that may contain \textit{only} static members.
For a C\# programmer, a static class is essentially a container for global variables and functions.
This is important since, in .NET, variables and functions cannot be declared outside of types.

\textbf{The abstract and sealed modifiers.}
Classes and their members may also be marked as either \texttt{abstract} or \texttt{sealed}.
Abstract classes cannot be instantiated, and all of their abstract members must be implemented by an inheriting class.
Sealed classes cannot be inherited from, and sealed type members cannot be overridden even if they were marked as abstract or virtual in a superclass.
Their function is mutually exclusive.

\textbf{The entity abstraction.}
Roslyn, NuGet, and MSBuild each have a different abstraction of the codebase they act upon.
For our purposes, we unify them into our \textit{entity abstraction}.
Each \textit{entity} has a name and an \textit{entity kind}, which can be, for example, \textit{solution}, \textit{project}, \textit{type}, or \textit{method}.
Entity kinds form a hierarchy since it can be said that a solution contains a project, which contains a type, which declares a method, and so on.
Some entity kinds are further specialized.
For example, types have a \textit{type kind} (e.g., class, struct, enum, etc.), and methods have a \textit{method kind} (e.g., constructor, getter, operator, etc.).

%-------------------------------------------------------------------------

\section{Visualization Design}
\label{sec:design}

Following Munzner's data visualization framework~\cite{munzner2014visualization}, we iterated and enhanced our design.
The main influences on the final visualization design and interactions were the \ref{req:first}--\ref{req:last} requirements, team discussions, and hands-on experience with our prototypes, including their previous evaluation~\cite{op}.

We considered several visual metaphors, including node-link diagrams, UML diagrams, and 3D cities.
Our goal was to create a visual representation that could provide both an overview of the entire project and details of individual code elements (see \ref{req:interact}).
We disregarded standardized UML diagrams because they would require different visual representations for the overview and the details, thereby increasing complexity and the learning curve, which was unacceptable given our aim to simplify documentation navigation.
Similarly, we dismissed the 3D city metaphor in favor of the simplicity and effectiveness offered by 2D diagrams.
Since the visualized data is not inherently spatial, we saw no justification for using a 3D visual channel~\cite{no_unjustified_3d}.

Therefore, we opted for a node-link diagram, representing the codebase as a graph, i.e.,~a set of nodes and edges.
Nodes correspond to \textit{C\# entities}, and edges represent relationships between them. Figure~\ref{fig:example} shows an example of this diagram.
While node-link diagrams were already used for visualization of codebases before~\cite{code_map, obsidian}, to the best of our knowledge, they were never explored as the central method for navigating data commonly accessible within an API reference.
We can also benefit from the fact that node-link diagrams are a well-known and generally understood concept, which enhances the adoption of the solution.

The rest of this section explains our visualization design from the smallest elements---glyphs---to the largest---the diagram's layout and user interface.
However, since the tool's implementation is not the main focus of this paper, we kindly refer the reader to our Supplementary Material~\cite{sm}.
In there, we detail the technologies, the user interface, and the system architecture with regard to requirements in Section~\ref{sec:requirements}.

\subsection{Glyphs}
\label{sec:glyphs}

% === New things to talk about ===
% [x] Custom icons
% [x] Node color presets
% [x] Filled icons for static
% [x] Donut charts
% [x] Contours for abstract and sealed
% [x] Collapsed node indicators
% [x] Diagnostic indicators

\begin{figure}[tb]
  \centering
  \includegraphics[width=0.58\linewidth]{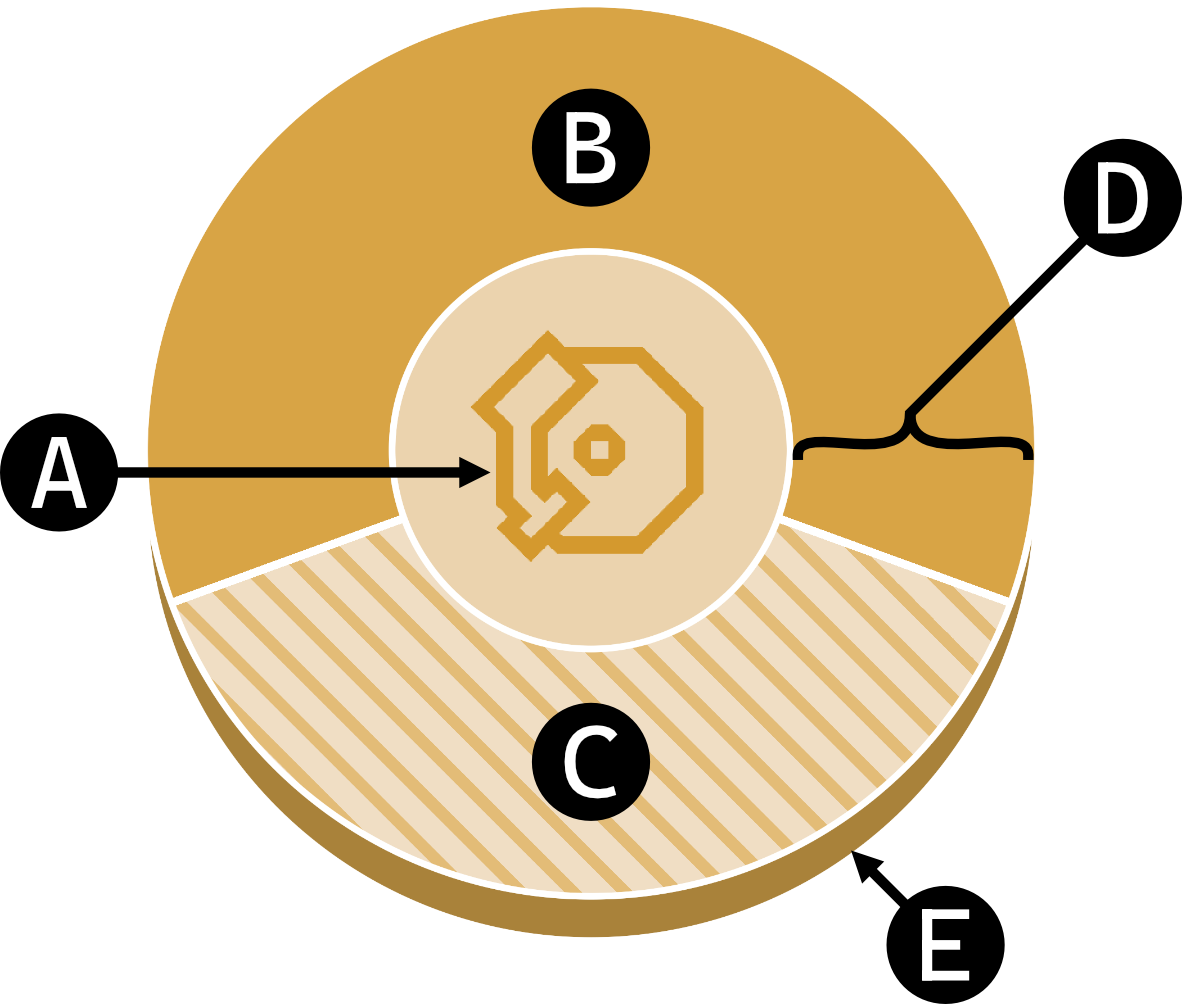}
  \caption{\label{fig:record-glyph} A glyph-node of a C\# type: (A)~An icon representing a \texttt{record class} type. The icon is stroked, implying the class is not \texttt{static}. (B~and~C)~Donut chart visualizing the ratio of static to instance type members, respectively. (D)~The width of the donut chart is proportional to the total number of type members. (E)~ The presence of a shadow below the nodes indicates that the node is collapsed and can be expanded.
}
\end{figure}

The nodes of the node-link diagram are glyphs---compound visual elements symbolizing several data attributes at once (see Figure~\ref{fig:record-glyph}).
We have decided to use glyphs as they provide a good compromise between space-efficiency and understandability~\cite{glyphs}.
All nodes are circular, as is typical for node-link diagrams.
Circles can also be arranged in a way that minimizes overlap and maximizes the use of space in the diagram, which is particularly useful for complex networks.

The node radius is calculated based on several factors.
First, the represented entity's kind determines the fixed starting radius.
For example, the starting radius of a project node is larger than that of a type node.
Nodes depicting the basic structure of the codebase (e.g., projects, namespaces, etc.) are then given additional size based on the height of their subtree, ensuring that, for example, nested namespaces are smaller than their parents.
Intuitively, this means that even among nodes of the same kind, hierarchy is visually represented.
For type entities, the starting radius is then increased by their member count.
The radius is then scaled, employing either a linear, logarithmic, or square-root scaling method.
Users have the flexibility to configure the scaling method, which can be helpful when the diagram contains excessively large nodes.

\begin{figure}[t]
  \centering
  \includegraphics[width=\linewidth]{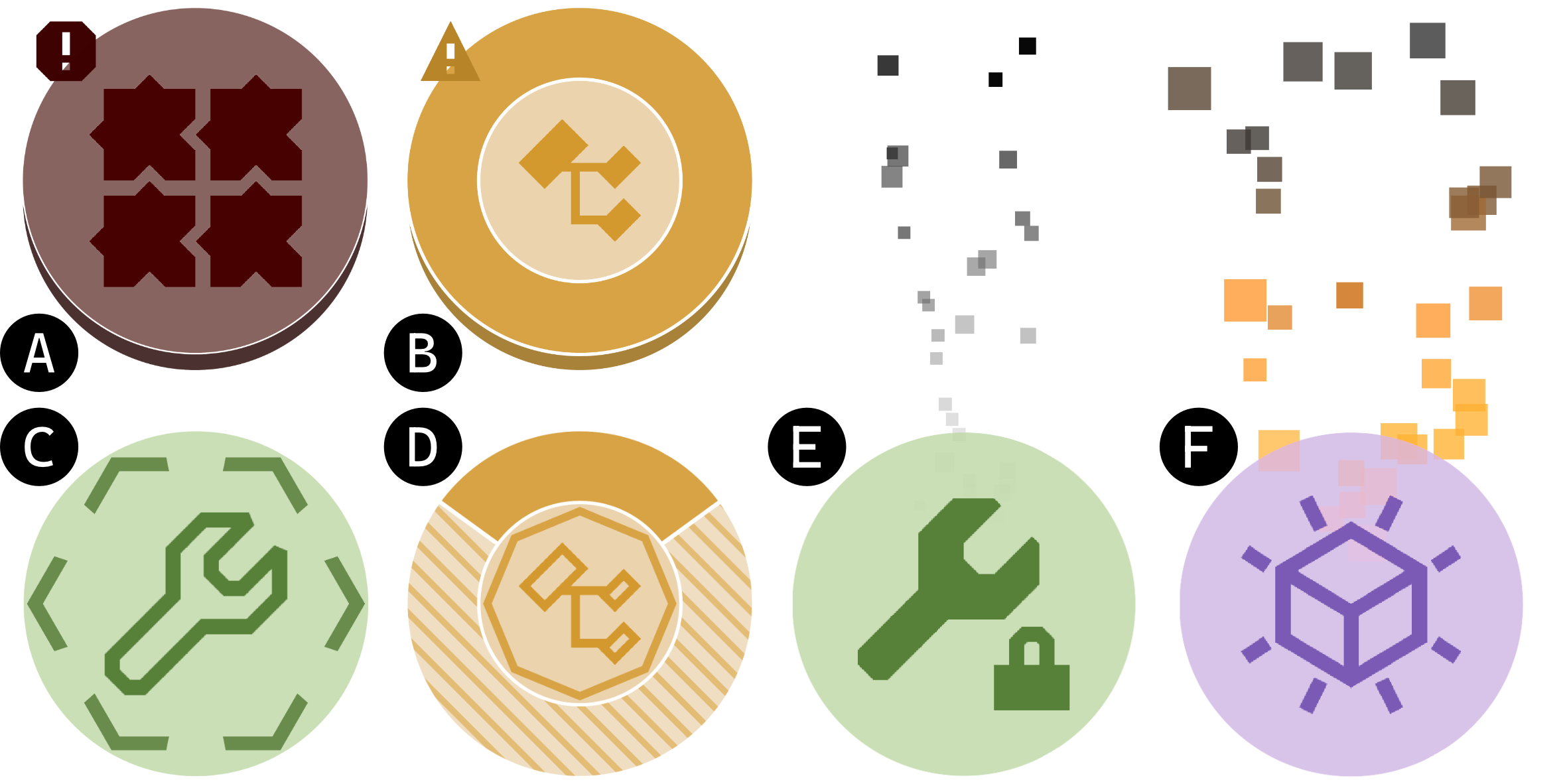}
  \caption{\label{fig:glyphs} Examples of glyph-nodes (rescaled): (A)~A~solution with an error in its subtree. (B)~A~static class with only static members and a warning in its subtree. (C)~An~abstract~property. (D)~A~sealed non-static class with both static and instance members. (E)~A~static private property with a warning. (F)~A~constructor method with an error.
}
\end{figure}

Each glyph has an icon in the middle representing its \textit{entity kind} (Figure~\ref{fig:record-glyph}-A).
In the case of types, the icon represents their \textit{type kind} as it is more specific (see Section~\ref{sec:csharp}).
Figure~\ref{fig:glyphs}-D, for instance, represents a \texttt{class} type, whereas Figure~\ref{fig:glyphs}-C is a property.
While in Helveg's previous version~\cite{op}, the icons were taken from the Visual Studio (VS) image library~\cite{vs_icons}, for this extended version, we opted to design our own icons and only take inspiration from VS, which aligns with~\ref{req:familiar}.
Our motivation for this decision was a glyph survey we conducted~\cite{op}, which showed that even programmers who used VS for many years have trouble recalling the meaning of the icons.
When given a test with 18 of the most relevant VS icons, the participants correctly identified their meanings only 68.52~\% of the time on average.
All of the new 98 icons are listed in Supplementary Material~\cite{sm}.
% Icons are from VS and reflect the kind.

In the past, the color of the nodes was determined by the VS icons.
While designing the new icons, we decided to make them monochromatic, allowing us to change the nodes' color scheme at any time.
However, this put before us the difficult task of creating a color palette with at least 17 distinctive colors, counting the most common entity kinds and all type kinds.
Ultimately, we included three node color presets in the new version: \textit{VS} is the old scheme for comparison, while \textit{Universal} and \textit{TypeFocus} are wholly new color palettes.
\textit{Universal} covers each entity kind with its own color, whereas \textit{TypeFocus} uses the most distinctive colors to differentiate only among type kinds while leaving most other entities a shade of gray.
Both new color presets were created manually with the help of \textit{iWantHue}~\cite{iwanthue} by Jacomy and \textit{chroma.js}~\cite{chroma} by Aisch.
To account for other use cases and color perception deficiencies, all colors are configurable from within the tool.

Another highly relevant information is \textit{accessibility} of types and their members. 
Its value can range from \texttt{public} to \texttt{private} with many nuanced shades in between~\cite{ecma-335}.
If the represented entity is anything other than \texttt{public}, the glyph contains a smaller icon in the lower right corner of the kind icon.
For example, Figure~\ref{fig:glyphs}-C is a \texttt{public} property since it has no icon in its corner.
However, Figure~\ref{fig:glyphs}-E is \texttt{private} because its corner icon depicts a lock.

As types are ubiquitous in C\#, they are essential in understanding a codebase.
Therefore, we apply several additional rules to types specifically.
Type icons exist in a \textit{stroked} and \textit{filled} variant.
You can see the difference, for example, in Figures \ref{fig:glyphs}-B and \ref{fig:glyphs}-D.
When a type icon is filled, the type is static and thus contains global variables and free functions.

All type nodes are also surrounded by a donut chart.
The chart always has up to two sectors, depicting the ratio of static to instance members of the type.
Similarly to the icon variants, the static member sector is filled, whereas the instance member sector features diagonal strokes (i.e., is cross-hatched).
In Figure~\ref{fig:record-glyph}, these are marked as B and C, respectively.
The width of the donut chart is proportional to the total number of type members, allowing for an intuitive comparison of types based on their size.
Previously, information about the static modifier and the type's member counts was carried by three concentric outlines around the icon~\cite{op}.
However, they proved counter-intuitive and difficult to read, with only  61.11~\% correct answers in the glyph survey~\cite{op}.
Therefore, they were replaced in the spirit of \ref{req:replace-outlines}.

In this improved version, we also newly visualize the \texttt{abstract} and \texttt{sealed} modifiers.
We use the fact that they are mutually exclusive and represent them by a polygonal ``contour'' around the icon.
We believe these visual elements to be fitting, as sealed entities cannot leave their octagonal enclosure (see Figure~\ref{fig:glyphs}-D), and abstract entities are incomplete just like the hexagon surrounding them (as in Figure~\ref{fig:glyphs}-C).

A subset of these rules also applies to type member nodes.
% A subset of these rules also applies to type members---nodes with the \textit{property}, \textit{field}, \textit{event}, or \textit{method} entity kind.
These nodes may also have accessibility icons, filled icon variants representing the \texttt{static} modifier, and polygonal contours for \texttt{abstract} or \texttt{sealed}.
However, they do not possess the donut chart, and their size is constant and smaller than that of any type node.

The glyph-nodes may also exhibit one of two animated effects.
These effects symbolize Roslyn diagnostics---compiler warnings and errors---which point out problematic pieces of code. They allow a team of developers to review issues within the codebase alongside the documentation.
Specifically, if a node has an error diagnostic attached to it, it appears to be on fire (Figure~\mbox{\ref{fig:glyphs}-F}).
In the case of compiler warnings, the node emits a mere column of smoke instead (Figure~\ref{fig:glyphs}-E).
These effects were chosen as fire and smoke are typically treated with caution, similar to compiler diagnostics.
We also chose to animate them, as that makes them more noticeable, which is fitting due to the diagnostics' importance.

Nodes with children may likewise possess \textit{indicators}, visual elements that inform about the node's descendants rather than the node itself (see \ref{req:indicators}).
% There are several types of indicators.
When the \textit{Collapsed node indicator} is present, as in Figure~\ref{fig:record-glyph}, it means the node has children, which are not currently visible and can be shown by expanding the node.
There are also \textit{Diagnostic indicators} announcing that one or more descendant nodes have compiler diagnostics.
These are shown in Figure~\ref{fig:glyphs} as little icons in the upper left corner of nodes A and B and symbolize the presence of errors and warnings in the subtree, respectively.

These rules are illustrated by the nodes shown in Figure~\ref{fig:glyphs}:
\begin{enumerate}[label={\helcircled{\Alph*}}, leftmargin=*]
    \item is the \textit{Solution} node and the root of the codebase. It also possesses an error diagnostic indicator meaning that at least one of its descendants has a compiler error.
    \item represents a \texttt{static} class with only \texttt{static} members.
        This is made clear by the filled icon and the presence of only the filled donut chart sector. It has a compiler warning in its subtree.
    \item is an \texttt{abstract} property, given by the property icon and the hexagonal incomplete (dashed) contour.
    \item is a non-\texttt{static} class with only a handful of static members. It is also \texttt{sealed} because its icon is enclosed in an octagonal contour.
    \item is a \texttt{static} property since it has a filled icon. It is \texttt{private} since it has a lock icon in its corner, and it has a compiler warning depicted by a column of smoke.
    \item is a constructor method since it has a specialized method icon. It also has a compiler error because it is on fire.
\end{enumerate}

In the context of glyph-based visualization research, our design follows several guidelines outlined by Borgo et al.~\cite{glyph_star} and Maguire et al.~\cite{glyph_taxonomy}.
The most important variable---\textit{entity kind}---is mapped to color, the channel with the greatest \textit{pop-out effect}~\cite{glyph_star}.
Entity kind is also redundantly mapped to glyph shape, as it determines the central icon.
While our glyphs have a complex composite shape that is not viewpoint-independent, in exchange, we are able to take inspiration from existing iconography.
Considering the abstract nature of the represented OOP concepts, we argue that familiar icons with metaphorical signs are more intuitive and memorable than abstract geometric shapes.
Another prominent channel, glyph size, is mapped onto the entity's position in the entity hierarchy.
Unfortunately, this transformation loses some information, as the absolute number of type members is not discernible.
However, since it is more important to recognize that a type is bigger relative to another type, we do not consider this a significant issue.
We also drew inspiration from the work of Maguire et al. on taxonomy-based glyph design~\cite{glyph_taxonomy}: Helveg's glyphs, and especially the central icons, follow a similar template.

\subsection{Entity Relationships}

Edges of the node-link diagram represent relationships between C\# entities.
Each edge is defined by its source and destination node as well as a \textit{relation}---a set of edges---it is part of.
There are multiple relations between nodes since we have decided to visualize more than just ``dependencies'' among the entities to represent the codebase accurately.

The \texttt{declares} relation mirrors the \textit{entity kind hierarchy} (see Section~\ref{sec:csharp}).
As such, it most clearly represents the codebase structure.
For example, in this relation, a \textit{solution} has outgoing edges into \textit{projects}, as solutions declare projects.
Similarly, a \textit{namespace} is linked to \textit{types} it contains, which are connected to \textit{methods} declared by them, and so on.
However, the \texttt{declares} relation alone offers no advantage over a table of contents commonly found in API references.
The real benefit of the node-link diagram lies in its ability to represent multiple relations at once.
Other relations symbolize entity associations such as type inheritance (\texttt{inheritsFrom}), types of fields/properties/parameters (\texttt{typeOf}), and project dependencies (\texttt{dependsOn}).
For a list of all relations that users can choose to visualize, see Supplementary Material~\cite{sm}.

To distinguish between relations when depicted simultaneously, each has a different edge color and thickness.
Each color is customizable, and each relation can be disabled if it is undesirable.
This way, the amount of clutter caused by an excessive amount of edges and their crossings can be minimized.
Following this mindset, we display only a subset of edges depicting the most important relationships by default.

\subsection{Scalability}
\label{sec:scalability}

A problem that node-link diagrams typically face when used to explore large datasets is visual clutter and overlapping, which can obscure important relationships and make the diagram difficult to interpret.
A common solution found in literature~\cite{schulz2011treevis} is to aggregate or filter the nodes. 
In this paper, we have primarily focused on filtering (see Section~\ref{sec:filtering}) as it was the most requested feature in the initial user survey~\cite{mgr}.
As we wanted to explore to what extent filtering would suffice as the primary means of improving the diagram's readability, we utilize only a simple form of aggregation by allowing users to expand and collapse nodes at will (see Section~\ref{sec:interactivity-layout}).

\subsection{Filtering and Highlighting}
\label{sec:filtering}

We strive to provide the most flexible filtering feature we can.
Therefore, we have decided on three search modes: \textit{full-text} search through the entity names, \textit{regex} doing the same but using a regular expression, and the \textit{JavaScript (JS)} mode allowing the user to write custom search logic with a JS filter.
In any mode, the user can either \textit{highlight} the matching nodes, graying out the rest, or \textit{isolate} the matches and remove the remainder from the current state of the diagram.

Although the JS search mode is the most flexible, in our previous study~\cite{op}, testers found it difficult to use, and some even outright refused to write JS code.
Thus, in the new version, we implemented the \textit{filter builder} component, which sacrifices some of the flexibility of the JS filtering for the sake of user comfort (in line with \ref{req:ergonomic-filtering}).
It allows users to pick node properties from a searchable list, choose one of the operators suitable for the property type, and filter by a specific value.
Only the nodes that fulfill all of the search criteria are included in the search results.
For example, the filter builder can be used to look up nodes with compiler errors, record class nodes, or entities with a certain keyword in their documentation comments.

\subsection{Interactivity and Layout}
\label{sec:interactivity-layout}

Apart from filtering, our approach relies on interactivity to be readable even as the diagram grows (see \ref{req:interact}).
By default, the diagram shows only the solution and project nodes, with the rest of the codebase hidden.
However, the user can choose a different starting graph by using the \textit{Quick Start} panel, such as displaying \textit{All types} of the codebase, \textit{Project dependencies}, or the \textit{Bird's eye view}, which displays nearly all nodes at once.
Then, they can use filtering and manual interaction to \textit{expand} nodes and show their children, thus shifting focus to a specific area of the codebase.
Likewise, nodes can be \textit{collapsed} to unclutter the diagram and gain a higher-level view.
Parts of the diagram can also be \textit{removed} from the current view.
These removed parts are then excluded from filtering and further interaction until the diagram is refreshed.
Finally, users can select any node they are interested in and inspect the represented C\# entity in a side panel.
Besides the entity's properties, documentation comments, and compiler diagnostics, the panel shows a preview of its node and the \textit{inspector}---the entity's C\# declaration, which we include to help users realize what they are currently inspecting.

The user interacts with the nodes using mouse and keyboard controls.
% Previously, all of these actions were bound to the left mouse button, and the user chose the current action by clicking on its icon in the \textit{Toolbox} component.
% This means of interaction was perceived as cumbersome and frustrating by participants of the previous user study~\cite{op}.
% Therefore, in this version, the Toolbox has been replaced with mouse and keyboard controls.
For example, nodes get expanded when double-clicked and can be manually moved by holding the shift button and dragging them with the mouse.
Each action also has a keyboard shortcut and is present in the context menu that appears when a node is right-clicked.
% We believe these new controls improve Helveg's intuitiveness and user experience.

The layout of the node-link diagram needs to respond to these interactions and remain readable.
% Hence, we decided to include an iterative force-directed layout algorithm.
Hence, we rely on two graph layout algorithms---TidyTree by Reingold and Tilford~\cite{tidytree}, and ForceAtlas2 by Jacomy et al.~\cite{forceatlas2}.
TidyTree positions the nodes in a circular dendrogram.
It executes almost instantaneously, but it requires the underlying relation to be a tree and thus is used only on \texttt{declares}.
ForceAtlas2 is an iterative, force-directed graph layout algorithm, which runs after TidyTree has been applied.
Compared to TidyTree, it is not limited to trees and is able to respond to user interaction, but its convergence may be slow, depending on the number of nodes.
Users can still reposition nodes manually if they are dissatisfied with the automatic layout.

In the updated version, the iterative layout algorithm is run after every node expansion or collapse and automatically stops when the average node traction~\cite{forceatlas2} drops below a configurable threshold.
We made this change based on feedback from the study participants: It is cumbersome to have to run the layout manually, but it is also important for the graph to cease moving so that it can be interacted with.

\begin{figure*}[tb]
  \centering
  \includegraphics[width=1.0\linewidth]{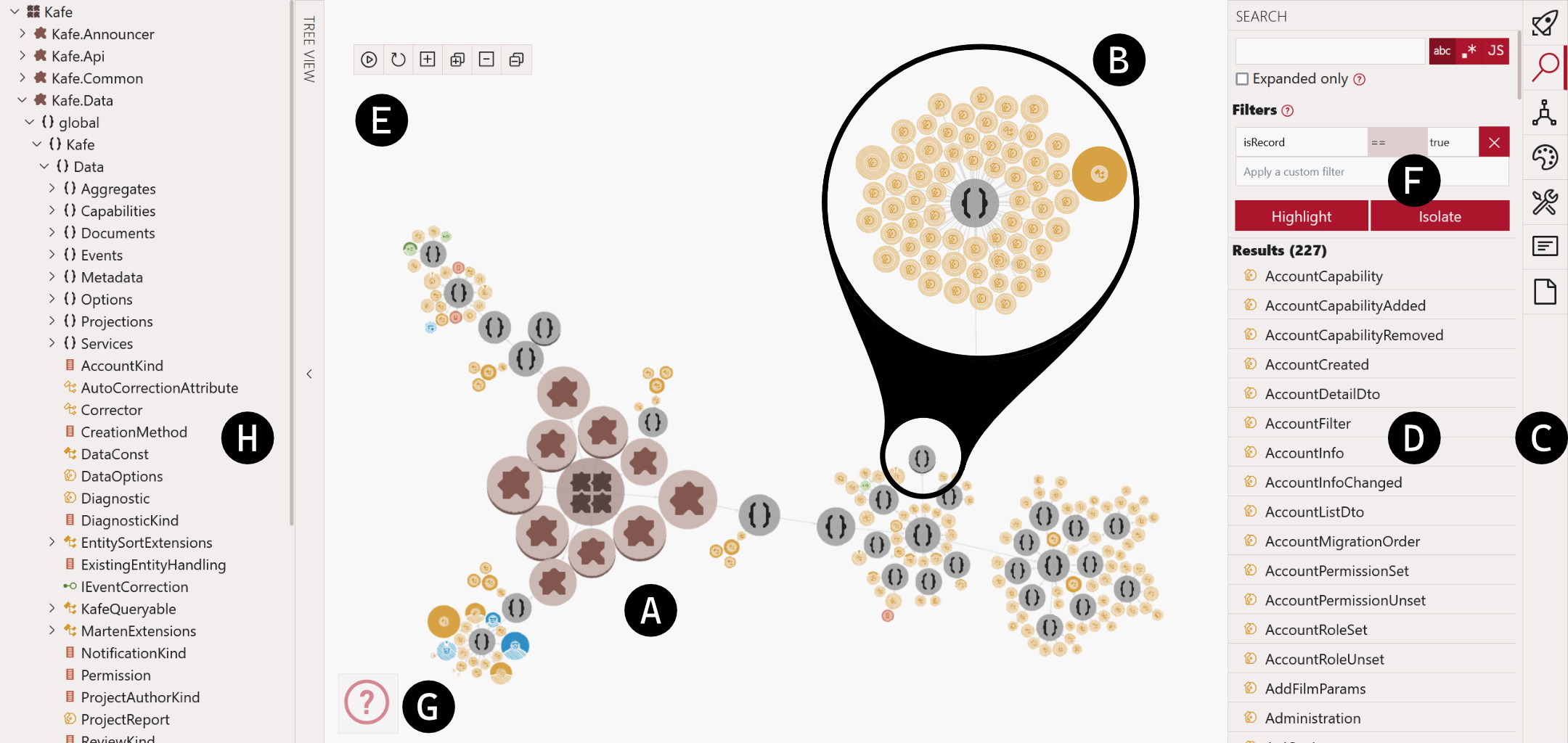}
  \caption{Helveg's user interface: (A)~The diagram with labels hidden for clarity. (B)~A node before (small circle) and after expansion (large circle). (C)~Dock allowing to switch between panels. (D)~Search panel. (E)~Toolbar with global actions. (F)~Filter builder. (G)~Cheat sheet button. (H)~Tree view.}
  \label{fig:ui} 
\end{figure*}

\subsection{User Interface}
\label{sec:user-interface}

The tool's user interface (UI) draws inspiration from existing software, such as graphical editors and Visual Studio Code, to be familiar to the user~(\ref{req:familiar}).
It consists of several distinct parts besides the diagram itself (Figure~\ref{fig:ui}-A).
The \textit{Dock} (Figure~\ref{fig:ui}-C) is a vertical bar housing several tabs.
It switches between panels, which occupy the same space (Figure~\ref{fig:ui}-D) on the screen.
Panels may contain the properties and documentation comments of the currently selected node, search results, or configuration options.
For example, in the \textit{Search} panel, there is the search bar and the filter builder component, acting as a more convenient way to build complex filters, as described in Section~\ref{sec:filtering}.
Actions that change the diagram globally (e.g., refresh or expand all nodes) are available through the \textit{Toolbar} (Figure~\ref{fig:ui}-E).
To the left of the Toolbar (Figure~\ref{fig:ui}-H) is the \textit{Tree View}, displaying the hierarchy of the codebase as a list, providing a view that is typical for an API reference.
Finally, in the lower left corner is the \textit{Cheat Sheet} button (Figure~\ref{fig:ui}-G).
When clicked, it displays an overview of all mouse and keyboard controls, glyph design rules, and panel contents.
It is complemented by a welcome screen, visible when the tool is first opened, and an interactive tutorial, which teaches the user about basic controls and points out all of the UI elements mentioned above.
To further ease users' onboarding, key UI elements contain hints and tooltips.

Most of the diagram's functionality can be customized to suit the user's needs.
All of the configuration options are contained in the panels.
For example, in the \textit{layout} panel, the user can use the provided checkboxes to select the entity kinds and relationships they are interested in.
Corresponding nodes and edges will be laid out once the \textit{Refresh} button is clicked.
The same panel houses controls of the layout algorithms.
All of the panels are pictured in Supplementary Material~\cite{sm}.

The new version makes several major alterations to the UI.
The \textit{Toolbar}, mouse controls, and keyboard shortcuts replace an earlier \textit{Toolbox}, which was considered as very cumbersome to use~\cite{op}.
The \textit{Cheat Sheet}, welcome screen, hints, and interactive tutorial supplant the previous version's \textit{Guide} panel, which was scarcely used by the original study participants.
Finally, the \textit{Tree View} has been added based on user feedback.

%-------------------------------------------------------------------------

\section{Helveg User Studies}

The Helveg prototype was thoroughly evaluated in two testing runs.
In the original paper~\cite{op}, we described the results of the first qualitative user study conducted with five software developers, who tested the prototype, and a glyph survey with six programmers that aimed to assess our glyph design.
For this extended and improved version of Helveg, we prepared and conducted another user study with the same developers as the first one (labeled as \textbf{E1}--\textbf{E5}) to ascertain that our changes addressed their original feedback.
We recruited these testers by word of mouth, and there was no reward for their participation in the study.
They have 7--12 years of professional C\# experience at a software company.
Only one of them suffers from a color perception deficiency---minor daltonism.
They have given their informed consent to the anonymized publication of their statements and experience.

% TODO: Talk about the first study
% [x] Observations: oversized ProjectService
% [x] Effects were very well received, that's why we don't test them again

The first user study featured tasks with no ``correct'' solution, which was a deliberate choice as this let us see how users approach the tool with little guidance.
The tasks were, however, chosen with documentation in mind and guided the participants towards getting a high-level overview of a codebase, which is one of the ways developers use API references~\cite{api_docs_usage}.
Among other things, we asked the participants to identify the purpose of each MSBuild project, their key external dependencies, and other symbols of interest.
While filtering and interactive exploration of the diagram, the testers made accurate observations about the codebase, its dependencies, and its purpose as a media archive.
They identified code elements deserving attention or refactoring, such as the oversized \texttt{ProjectService} node.
While their overall feedback was positive, especially when compared to existing tools, they were dissatisfied with the user interface and cumbersome controls, which reflected poorly in their rating of the tool's interactivity and intuitiveness (see Figure~\ref{fig:results}).

The first study was followed by a glyph survey, which tested how well C\# programmers recognize icons from VS and whether our old glyph design was readable.
The participants of this survey were six developers, one of whom also previously tested the prototype.
The results were less successful than we expected, with only 68.52~\% correctly assigned icons on average, and only 61.11~\% correct readings of our glyphs with \textit{outlines}~\cite{op}.
However, all of the participants \textit{Agreed} or \textit{Strongly agreed} that the animated effects were a faithful representation of compiler diagnostics.
Therefore, in our new glyph design, we chose to keep these effects as is and exclude them from the second study.

The second user study focused solely on assessing new features and their efficiency.
Throughout testing, the participants performed specific tasks and filled in a questionnaire.
While carrying out the tasks, they worked with a sample visualization of the KAFE project~\cite{kafe}, an open-source web archive of multimedia.
They had no experience with the project, besides it being a sample used in the previous user study.
However, the interviewer was one of KAFE's developers and thus could easily access the testers' observations.
The decision to keep the same project for the study was intentional, as this time, we wanted to validate the design improvements and not the understanding of the project content. 
A summary of each session, the questionnaire, the testers' answers, and the KAFE sample are a part of Supplementary Material~\cite{sm}.

\subsection{Testing Sessions}

The second study's testing sessions consisted of multiple parts.
First, the participants answered several general questions about themselves.
Then, they were asked to perform seven tasks, focusing solely on the new features.
After each task, they were asked an open question and one or more questions on a five-point Likert scale.
At the end of the \textit{Testing} section, they shared their experience and rated the prototype with respect to aspects such as readability, interactivity, and UI design.
Finally, they were given several questions regarding our glyph design, which were similar to our previous glyph survey~\cite{op}.
The testing sessions took between 58 to 96 minutes, based on each tester's individual speed and time constraints.

% [x] Tutorial and Cheat Sheet
% [x] Mouse and keyboard controls
% [x] Filter builder
% [x] Other UI: Tree view, node preview, and inspector
% [x] Glyphs: appearance, icons, indicators

First, they were asked to complete the interactive tutorial and read the cheat sheet.
These features were meant to address comments the participants made in the first study~\cite{op} about the overwhelming amount of options.
The tutorial and cheat sheet were accepted very positively, with all of the participants \textit{Strongly agreeing} that they are an improvement over the previous version's Guide panel, which contained only a short description of the tool.
They also \textit{Agreed} or \textit{Strongly agreed} that hints complemented the tutorial and cheat sheet well.
However, one tester suggested to \textquote[\textbf{E3}]{Show, don't tell.} saying the tutorial could be even more interactive.

With regards to the new mouse and keyboard controls, all testers felt it was definitely an improvement over the Toolbox, with four testers \textit{Strongly agreeing} and one \textit{Agreeing}.
However, they disliked the slowness of the graph layout with feedback such as \textquote[\textbf{E1}]{The nodes moving beneath my hands is inconvenient.}
They also expressed this opinion when asked whether the auto-stopping behavior is sufficient, since two \textit{Disagreed}, two \textit{Agreed}, and one was \textit{Undecided}.

When it came to the filter builder, all unanimously \textit{Strongly agreed} it is an improvement over the previous search methods, with comments such as \textquote[\textbf{E2}]{It was very easy to use.}
They were also fond of the component's automatic completion feature.
Still, they offered several improvements, like adding a selection component for node properties with a limited domain of values such as C\# accessibility modifiers (see Section~\ref{sec:csharp}).

Similarly, all testers either \textit{Agreed} or \textit{Strongly agreed} that the other newly added UI elements, namely the hints, the Tree View, and the node preview and inspector in the Properties panel, were useful.
They did suggest, however, that \textquote[\textbf{E1}]{There could be a checkbox like `Sync solution with current item' in VS.}, which would automatically expand elements in the Tree View based on the currently selected node.

Several questions focused on the new glyph design.
The icons were praised with two \textit{Strongly agreeing} and three \textit{Agreeing} participants that they were an improvement over their predecessors.
They also felt very positively about the change from outlines to donut charts and icon variants and about the inclusion of the contours for the \texttt{abstract} and \texttt{sealed} modifiers.
\textquote[\textbf{E1}]{Nicely recognizable.} and \textquote[\textbf{E4}]{They’re very intuitive.}, some commented.
All but one tester \textit{Agreed} that nodes that can be expanded are clearly marked by their shadow.
The one who \textit{Disagreed}, however, said \textquote[\textbf{E1}]{I don't know how else to mark it.}
Interestingly, the \texttt{record class} icon resembling a vinyl record was especially controversial, with two participants saying \textquote[\textbf{E1}]{I needed to get accustomed with [the] record icon} and \textquote[\textbf{E3}]{Record [icons] are strange.}, while others said statements like \textquote[\textbf{E4}]{I very much like that they look like LPs. Very wholesome.} 
Even more divisive was their preference of style for the instance member sector of the donut charts.
Two felt the cross-hatching helped it be distinguishable and \textquote[\textbf{E4}]{visually pleasing}, while two felt it was \textquote[\textbf{E3}]{visually noisy}, with one remaining \textit{Undecided}.

Throughout testing, the participants suggested many minor user experience improvements, which we omit in this paper for brevity.
However, they are all included in the interview notes in Supplementary Material~\cite{sm}.

\subsection{Results}
\label{sec:helveg-results}

% [x] Open feedback
% [x] Ratings
% [x] Glyph retention

In both studies, after completing the tasks, each participant was asked what they enjoyed and what should be further improved. In the first study, the tool's visualization was well-received and considered readable and useful, as one of the participants remarked:

\begin{displayquote}[\textbf{E5}]
I enjoyed seeing the graph and the fact that I could imagine what the [VS] solution looks like without looking at any piece of code.
\end{displayquote}

When asked to compare Helveg to existing tools, the participants responded positively.
One developer considered Helveg to be well-performant, saying that \textquote[\textbf{E2}]{Code Map in Visual Studio usually takes so long.}
Another tester said that \textquote[\textbf{E4}]{For the amount of information that is shown right here, the JetBrains UML diagram would be unreadable for me.}
Some saw it as a helpful onboarding tool, and even the most critical tester approved of the tool's basic concept: \textquote[\textbf{E3}]{I like the idea itself. I like it for the API reference, especially.}

However, the testers expressed dissatisfaction with the intuitiveness of the original prototype's UI. Multiple participants also disliked that the most flexible way to filter the graph was through JavaScript, a language they detested. We addressed both of these criticisms in the tool's current iteration.

All participants of the second study considered the new version to be a considerable improvement.
They enjoyed the new filtering method, the Tree View, the new glyph design, and the mouse controls, with one participant going as far as to say:

\begin{displayquote}[\textbf{E5}]
The jump from the old version to the new is a big one. I enjoyed seeing the changes made. With [a] much cleaner user interface, tutorial, and helping tooltips for nearly everything the user is trying to do, it’s a very
usable tool.
\end{displayquote}

On the other hand, the most frequently mentioned features deserving improvement were the speed of the layout, the extent of data mining (e.g., to constant values and maintainability metrics), and minor improvements to Tree View (e.g., synchronization with the diagram).
Notably, the participant suffering from a color perception deficiency, expressed no discomfort or difficulty using the tool.

\begin{figure}[tb]
  \centering
  \includegraphics[width=\linewidth]{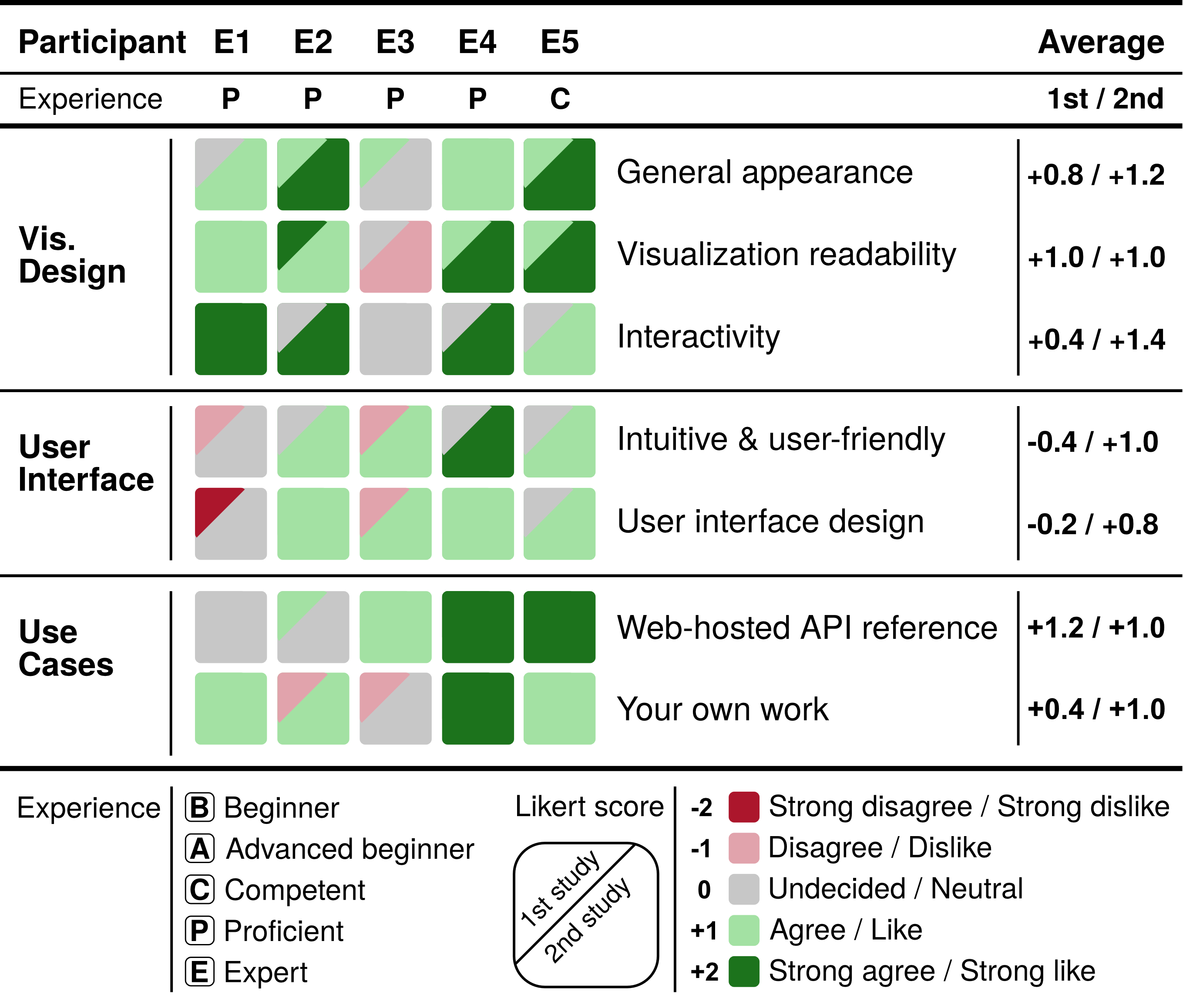}
  \caption{\label{fig:results} Questionnaire results from both user studies with color-coded Likert scores and an average result for each question.}
\end{figure}

Like in the previous study, the participants rated their experience with Helveg in several areas on a five-point Likert scale (see results in Figure~\ref{fig:results}).
Most notably, the intuitiveness, interactivity, and user interface of the tool saw strong improvement, likely largely thanks to the filter builder, since the testers previously heavily disliked the filtering experience.
While some considered the visualization readability to improve as well, probably thanks to the new glyph design and color presets, one tester (\textbf{E3}) decided to give it a lower score, quoting the slowness of the layout as the reason.
The rest of the scores did not see a strong change either way.

The last section of the questionnaire tested whether the participants recalled the iconography and glyph rules after using the tool for a short time.
First, they were given an image of twelve icons and asked to assign them a name from a predefined set.
All testers were able to correctly assign the majority (93.33~\% on average) of them, with the only two mistakes being the swapping of either methods and properties or fields and method parameters.
Then, they were given an image with a mixture of type and type member glyphs and a series of eight questions about the entities they represent.
Again, all testers answered the majority of the questions correctly (95~\% on average).
While the initial glyph survey was conducted with a different group of participants and under different circumstances, this result shows a great improvement in the glyphs' readability.

\subsection{Threats to Validity}

While we made every effort to evaluate Helveg fairly, some threats to the evaluation's validity remain.
Both studies dealt with a small pool of developers ($n = 5$).
While this allowed us to conduct our user studies qualitatively and discuss each tester's experiences in detail, it remains to be seen whether the tool would be accepted by the developer community at large.
Our testers were also experienced developers, and we cannot be certain that our approach would be as suitable for novices.
Another limitation is the focus of the studies on exploring how programmers use our tool.
Although this highlighted the advantages and helped us discover the flaws in our approach, a drawback of this decision is that neither study featured tasks that would allow us to draw direct comparisons between Helveg and a traditional API reference.
Finally, despite our desire to test each icon set under the same circumstances, there was a difference between the studies, which arose naturally because of the different origins of each icon set.
In the first study, we tested whether C\# developers recall Visual Studio icons they have been seeing over their whole career, whereas in the second study, they assigned meanings to our custom icons, which, although similar to the first set, they had seen for the first time during the testing session.

%-------------------------------------------------------------------------

\section{Discussion}

Both user studies with C\# programmers and the glyph evaluation show that our diagram can provide a high-level overview of a codebase.
They also suggest that our approach, and especially our prototype, still suffers from several issues.
In this section, we discuss the lessons learned, how we addressed some of them in the new version, and what can be done to address the rest of them in the future.

\subsection{Lessons Learned}

\vspace{0.5em}\noindent\fbox{\begin{minipage}{0.98\linewidth}
\vspace{0.1em}
    Our glyphs and diagram-based navigation can be used to gain knowledge about a codebase that a typical API reference can hardly provide.

\vspace{0.1em}
\end{minipage}}\vspace{0.5em}

The testers' observations show what can be learned from the diagram.
For example, their discovery of the need to refactor the oversized \texttt{ProjectService} class could be made with a typical API reference only with great difficulty.
In the API reference, the number of type members is not discernible unless one reads every type's page.
Another example is the visualization of compiler diagnostics since, to the best of our knowledge, no API reference generator includes those in its output.
In order to see those, the reader would need to download the source code, which may be unavailable, and compile it themselves, which could require a significant amount of time.

\vspace{1em}\noindent\fbox{\begin{minipage}{0.98\linewidth}
\vspace{0.1em}
    A familiar interface and a set of icons do not necessarily make a tool intuitive.
    A tutorial and other hints are essential.
    
\vspace{0.1em}
\end{minipage}}\vspace{0.5em}

While we did our best to make the UI of the first Helveg version familiar, it did not make up for a lack of a proper tutorial and other learning features.
A single short description of the tool's features tucked away in a panel has proven insufficient.
The comments the testers made and the difference in the perception of the tool's intuitiveness between the two versions show that features, such as an interactive tutorial, cheat sheet, and hints, are essential for a tool as configurable as Helveg.
Without these features, users may get overwhelmed and even discouraged from using the tool.

% We also saw improvement in the readability of our glyphs.
% This is in part due to the replacement of the problematic visual element of outlines with more common donut charts, as the testers and their results on the glyph retention questions concur.

\vspace{1em}\noindent\fbox{\begin{minipage}{0.98\linewidth}
\vspace{0.1em}
    A complex composite glyph can cover many variables and their values.
    Taking inspiration from existing domain-specific iconography is advisable.
    
\vspace{0.1em}
\end{minipage}}\vspace{0.5em}

Our glyphs can serve as an example to the broader visualization community on how to design domain-specific glyphs and what to avoid.
We created icons that cover a large space of values in a similar way to Maguire et al~\cite{glyph_taxonomy}.
Specifically, we mapped most of .NET's type system and its many modifiers to visual elements such as: icons in two different styles (\textit{stroked} and \textit{filled}), smaller icons in corners of the glyphs, contours, and visual metaphors (e.g., a vinyl record for the \texttt{record} type).
We also recommend including meta-variables in glyph design.
For example, Helveg shows whether a node is collapsed or expanded by rendering a shadow beneath collapsed nodes, suggesting there is more to be explored.

We also echo a guideline of Borgo et al.~\cite{glyph_star}: Avoid mapping a variable onto the radius of a circle.
We have demonstrated this through \textit{outlines}---mapping of the number of \texttt{static} and non-\texttt{static} type members onto the radii of two concentric circles.
In the second evaluation, we replaced these with more common donut charts and saw improvement in the readability of our glyphs.
Readability has also improved thanks to the new icons.
The previous version featured icons taken directly from the Visual Studio Image Library~\cite{vs_icons}, whereas the new icon set is only inspired by VS.
Compared to VS icons, they are simpler, cover more special cases, and fix some confusing design choices we noticed during our glyph survey~\cite{op}---such as with the VS icon for \textit{fields} and \textit{method parameters} being almost identical.
Our icons are also monochromatic, allowing us to create different color schemes, such as \textit{Universal} and \textit{TypeFocus}.
Consequently, we believe it to be a good practice to exploit the familiarity of pre-existing domain-specific icons but to use them only as a basis for custom visual elements.

The glyph effects received an overwhelmingly positive response in our glyph survey~\cite{op}.
Although the survey's sample size was small, this result indicates that the effects are suitable visual representations of compiler diagnostics and possibly similar phenomena in other domains.

\vspace{1em}\noindent\fbox{\begin{minipage}{0.98\linewidth}
\vspace{0.1em}
    A filtering feature must balance its flexibility with its ease of use.
    Even programmers may prefer not to program.
\vspace{0.1em}
\end{minipage}}\vspace{0.5em}

We also learned that simple features are not always easy to use.
The JavaScript filter mode, which we implemented in Helveg, is simple in its implementation and flexible in its use cases.
However, C\# developers were reluctant and felt uncomfortable writing code in JS.
The addition of the filter builder component in the new version solved this issue, allowing users to construct filters more ergonomically at the expense of some loss of flexibility.
We believe this finding applies to any visualization using a filterable node-link diagram, especially those targeting users other than programmers.
Nevertheless, in the second user study, testers still found room for improvement.
For example, the filter builder currently offers no way of importing and exporting a query, which are features that could be helpful in a collaborative environment.

\vspace{1em}\noindent\fbox{\begin{minipage}{0.98\linewidth}
\vspace{0.1em}
    The prototype neither scales well nor performs well when applied to large codebases.

\vspace{0.1em}
\end{minipage}}\vspace{0.5em}

Although, according to the testers, Helveg performs better than existing tools, such as Visual Studio Code Map and the JetBrains UML plugin, it does not scale well on projects much larger than KAFE.
KAFE comprises eight projects, resulting in a diagram with at most several thousand nodes when completely unfiltered and expanded.
The issue is further exacerbated with the number of relations shown at once.
However, C\# solutions can grow to hundreds of projects and tens of thousands of nodes.
We knew about this scalability issue when we designed the visualization (see Section~\ref{sec:scalability}), so we included flexible filtering features, allowing the user to choose what to focus on.
Without the filtering features, however, the prototype currently cannot handle large solutions.

When considering the requirements for the tool's new version (see Section~\ref{sec:requirements}), we have decided to leave the layout unchanged.
This decision allowed us to test the new glyph design and other changes without the uncertain impact of a completely different layout algorithm.
However, it also meant that the scalability issue remains.
Based on tester feedback, the layout is currently Helveg's largest outstanding problem, which impairs the visualization's readability and interactivity.

In the future, we may consider alternatives to the node-link diagram, such as a circular treemap or a hybrid of the two techniques designed for large codebase exploration.
Helveg's performance bottleneck can likely be solved by rendering the visualization using WebGPU~\cite{webgpu}, a future web standard that may be able to visualize a much greater number of nodes.

\vspace{1em}\noindent\fbox{\begin{minipage}{0.98\linewidth}
\vspace{0.1em}
    Configuration is vital due to varying personal preferences.

\vspace{0.1em}
\end{minipage}}\vspace{0.5em}

In the new user study, we saw that icon styles and contours are a welcome addition to the glyph design rules.
We also found that changing the node color palette helps cover multiple use cases, such as focusing on the .NET type system with \textit{TypeFocus}.
However, as the question about cross-hatching has shown, preferences and color vision differ among users.
Thus, we emphasize the need for options to turn these visual elements off and to modify color palettes.

\subsection{Future Work}

% Maybe we don't need the diagram to be so interactive. We have observed certain patterns in what programmers tend to look for. Maybe it would be worth trying to precompute for those use cases first --- optimize for the hot paths if you will --- and let them compute their obscure use cases on demand, albeit slowly.

There are many ways to extend our approach besides those implied by the already discussed lessons learned.
The most obvious extension would be to support another programming language.
Another potential extension can be found in the visualization of differences between two codebases or rather two versions of the same codebase.
This could be especially useful for code reviews.
The data miner can be extended as well.
For example, it could analyze method bodies, leading to a diagram depicting the method call tree, or provide more details about package references, such as whether they are direct or transitive.
The usability of the tool could also benefit from tighter integration into a code editor or a developer web platform (e.g., GitHub).
However, this integration would bring its own limitations since the developer community is fractured in the editors and platforms it uses~\cite{so_survey}, and it would take great effort to support them all.
Besides new features, a user study with more participants with diverse experience levels should be done to test this approach.

%-------------------------------------------------------------------------

\section{Conclusion}
    
We introduced a visual approach for automatically-generated API documentation, where navigation is handled by a central node-link diagram.
It provides a high-level view with interactive features, allowing in-depth exploratory analysis of a codebase.
The diagram's glyph-nodes take inspiration from iconography that most users already know, while also simplifying and extending it.

We created Helveg, a prototype implementation of this approach.
Through repeated evaluation with professional software developers, we conclude that Helveg is functional and can provide insight into a codebase, which would be hard to achieve using a typical API reference.
In this extended version of our paper, we also addressed issues in our glyph design and the tool's intuitiveness and interactivity.

%-------------------------------------------------------------------------

\section*{Acknowledgements}

We thank all of our testers for their continued interest in our tool.
We also thank Nika Kunzová for her invaluable feedback on Helveg's user interface and icon design.

%-------------------------------------------------------------------------

\bibliographystyle{IEEEtranDOI}
\bibliography{IEEEabrv,paper}

\vspace{-30pt}
\begin{IEEEbiography}[{\includegraphics[width=1in,height=1.25in,clip,keepaspectratio]{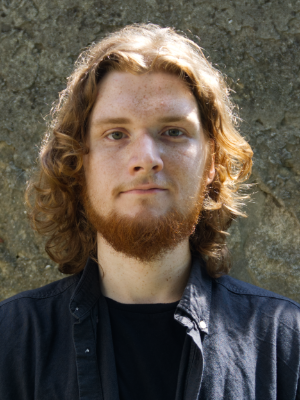}}]
{Adam Štěpánek} is a PhD student specializing in software visualization at the Visitlab research laboratory at the Faculty of Informatics, Masaryk University in Brno, Czech Republic.
His topic involves mining data related to software and visualizing said data in a preferably playful way.
\end{IEEEbiography}
\vspace{1pt}
\begin{IEEEbiography}[{\includegraphics[width=1in,height=1.25in,clip,keepaspectratio]{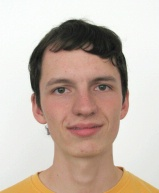}}]
{David Kuťák} is a PhD student at Visitlab, Faculty of Informatics, Masaryk University in Brno, Czech Republic. After receiving his Master’s degree (Mgr.) in 2019 from Masaryk University in Brno, Czech Republic, he continued a computer science path. His research topic involves biochemical visualization, focusing mainly on DNA and proteins, combined with Virtual Reality environments.
\end{IEEEbiography}
\vspace{1pt}
\begin{IEEEbiography}[\href{https://www.fi.muni.cz/~xkozlik/}{\includegraphics[width=1in,height=1.25in,clip,keepaspectratio]{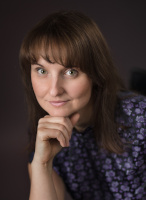}}]{Barbora Kozlíková} is an Associate Professor at the Faculty of Informatics at Masaryk University in Brno, Czech Republic. She is the head of the Visitlab research laboratory, specializing in the design of visualization and visual analysis methods and systems for diverse application fields, including biochemistry, medicine, and geography. She has published over 100 research papers.
\end{IEEEbiography}
\vspace{1pt}
\begin{IEEEbiography}[{\includegraphics[width=1in,height=1.25in,clip,keepaspectratio]{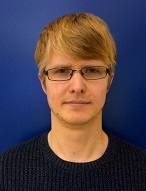}}]{Jan Byška}is an Assistant Professor at the Masaryk University in Brno, Czech Republic and a part-time Associate Professor at the University of Bergen, Norway. He is a member of the Visitlab research laboratory, where his work focuses mostly on various challenges in the field of visualization of molecular and time-dependent data.\end{IEEEbiography}

\end{document}